\documentclass[aps,prl,twocolumn,groupedaddress]{revtex4-2}
\usepackage{graphics,epsfig} 
\usepackage{xcolor}
\begin{document}
\title{Stochastically perturbed billiards: fingerprints of chaos and universality classes}
\author{Roberto Artuso$^{1,2}$}
\email{roberto.artuso@uninsubria.it}
\author{Matteo Burlo$^{1}$}
\email{matteo.burlo46@gmail.com}
\affiliation{$^{1}$Center for Nonlinear and Complex Systems and
  Dipartimento di Scienza e Alta Tecnologia, Via Valleggio 11, 22100
  Como (Italy)} 
\affiliation{$^{2}$I.N.F.N., Sezione di Milano, Via Celoria 16, 20133
  Milano (Italy)} 
\date{\today}
\begin{abstract}
Billiards tables - a minimal model for particles moving in a confined region - are known to present classical (and quantum) different features according to their shape, ranging from strongly chaotic to integrable dynamics. Here we consider the role of a stochastic perturbation of the elastic reflection law, and show that while chaotic billiards maintain their key statistical feature, the behaviour for integrable billiard tables is completely different: it can be linked, for tiny perturbations, to Evans stochastic billiard, where at each collision the reflected angle is a uniformly distributed stochastic variable on $(-\pi/2,\pi/2$). The resulting spatial stationary measure has peculiar aspects, like being typically non uniform along the boundary, differently from any chaotic billiard table.
\end{abstract}
\maketitle
\begin{figure}[h!]
\begin{center}
\includegraphics[width=7.4cm]{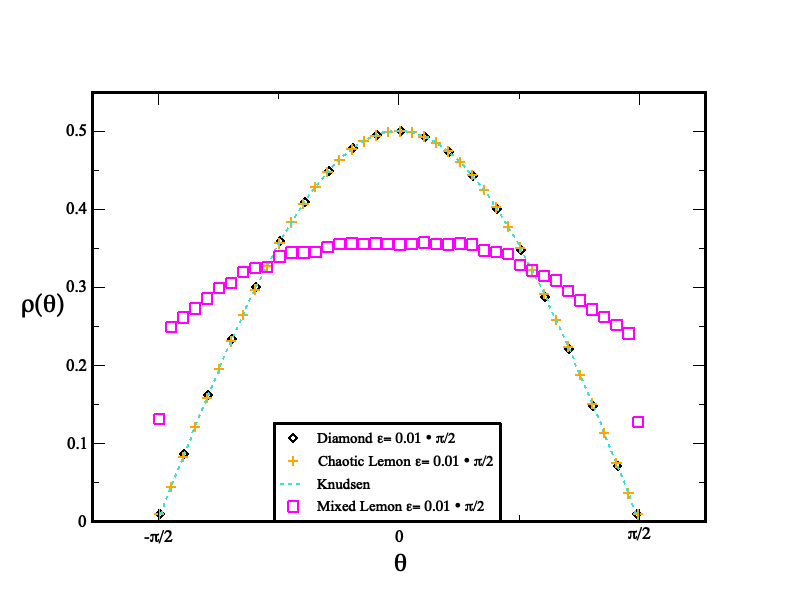}
\end{center}
\caption{(color online) {Probability density of outgoing angles under a weak stochastic perturbation: Knudsen distribution is well reproduced by fully chaotic diamond billiard ($R=\sqrt{2}$ and $L=1+\sqrt{3}$) and ergodic lemon billiard ($R_1=1,\,R_2=1.5,\,d=1.15$), while significant deviations appear for $R_1=R_2=1,\,d=0.01$, close to the integrable case (circular billiard). All numerical data are obtained by $10^9$ collisions.}}
\label{dia-a}
\end{figure}
\begin{figure}[h!]
\begin{center}
\includegraphics[width=7.4cm]{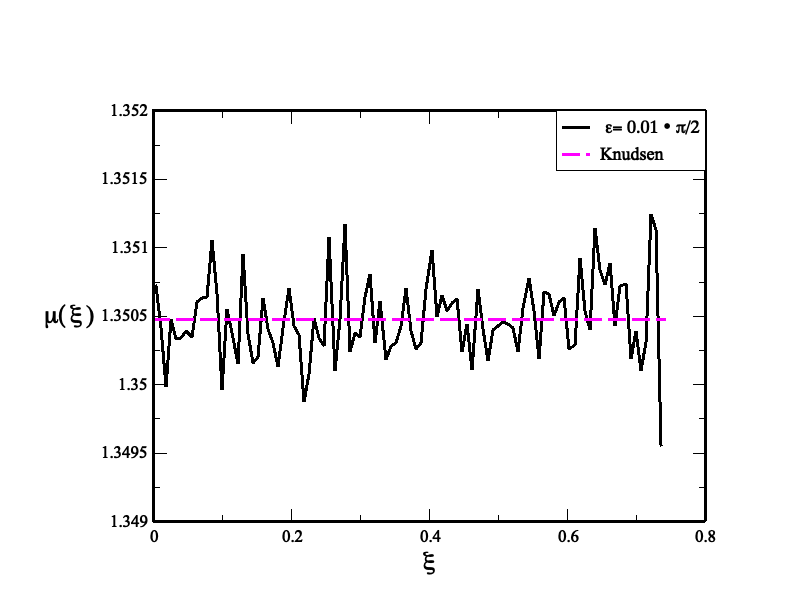}
\end{center}
\caption{(color online) {Probability density along the boundary under a weak stochastic perturbation for the chaotic diamond billiard ($R=\sqrt{2}$ and $L=1+\sqrt{3}$): we observe small fluctuations around a uniform density (due to symmetry only one out the four arcs is considered). Numerical data are obtained by $10^9$ collisions.}}
\label{dia-s}
\end{figure}
\begin{figure}[h!]
\begin{center}
\includegraphics[width=7.4cm]{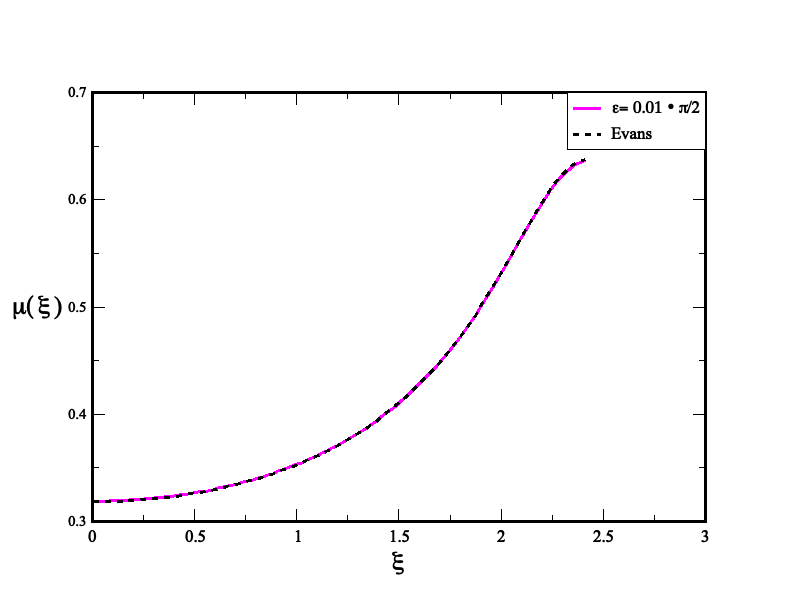}
\end{center}
\caption{(color online) {Probability density along the boundary for one quarter of an elliptical billiard ($a=2,\,b=1$) under a weak stochastic perturbation: numerical data (with $10^9$ collisions) are compared to the invariant density of the Evans billiard, eq. (\ref{invEdb}).}}
\label{ell-s}
\end{figure}
\begin{figure}[h!]
\begin{center}
\includegraphics[width=7.4cm]{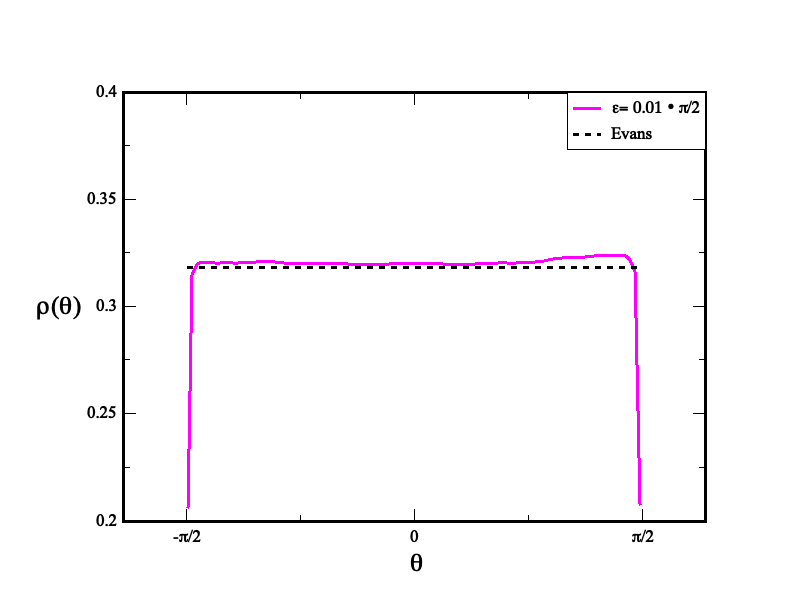}
\end{center} 
\caption {(color online) {Probability density of outgoing angle for one quarter of an elliptical billiard ($a=2,\,b=1$) under a weak stochastic perturbation: numerical data (with $10^9$ collisions) are compared to a uniform distribution: deviations only appear close to the boundary (see text).}}
\label{ell-a}
\end{figure}

Billiard tables, where a particle moves with constant velocity in a bounded domain, being elastically reflected upon colliding with the boundary, are known to display a rich variety of dynamical behaviour \cite{Tab,ChMa,KH}, dictated by the geometric shape of the boundary. In particular dispersing and defocusing billiards \cite{bun-non,PBill} are prototypical examples of ergodic and chaotic systems, while, on the opposite side, circular or elliptic billiards provide instances of regular, integrable motion \cite{KaSo}. We recall that this classical properties also reflect at the quantum and mesoscopic level \cite{oriol,fritz,NaHa}.
\\
Elastic reflections are defined by the velocity change rule at the point of impact $\mathbf{q}$:
\begin{equation}
\label{el-coll}
\mathbf{v}_{out}=\mathbf{v}_{in}-2\mathbf{n}_q (\mathbf{v}_{in}\cdot \mathbf{n}_q),
\end{equation}
where $\mathbf{n}_{\mathbf{q}} $ is the (inward) normal at $\mathbf{q}$. The law (\ref{el-coll})  conserves the kinetic energy of the particle, and it guarantees that phase space volume is preserved \cite{ChMa} (so the dynamics is Hamiltonian). We mention that {\it deterministic} reflection laws different from specular reflections have been considered to study shear flows, in connection with fluctuation relations of phase space contraction \cite{ChLe,BCL}.\\
On a different perspective, stochastic reflection laws have been considered since the foundation of kinetic theory of rarefied gases \cite{KG}: in particular by taking into account the Knudsen (or Lambert \cite{Chand}) cosine law, where, upon collisions, the outgoing angles $\psi_i$ (measured w.r.t. $\mathbf{n}_q$) are i.i.d. random variables, independent on the incoming angle $\phi_i$ and distributed with a density
\begin{equation}
\label{KnudLamb}
p_{K}(\psi)=\frac 1 2 \cos(\psi) \qquad \psi \in (-\pi/2,\pi/2).
\end{equation}
Later Knudsen prescription has been modified \cite{Smol,rou1}, considering a mixture of elastic and Knudsen collision law, and this picture was supported by microscopic simulations \cite{KCel}. \\
While stochastic Knudsen law (\ref{KnudLamb}) has been mathematically supported as a model of collisions from rough surfaces \cite{FY,Feres-r}, and it enjoys nice statistical properties \cite{CPSV}, other probability distributions have been analysed: in \cite{Evans} billiard tables with a uniform scattering angle (again independent of the incoming angle), corresponding to a density
\begin{equation}
\label{pEvans}
p_{E}(\psi)=\frac 1 {\pi} \qquad \psi \in (-\pi/2,\pi/2),
\end{equation}
were introduced. Evans proved that, both for continuous time dynamics, and the corresponding collision Markov chain, there exists a unique invariant probability measure. Such a measure displays peculiar properties: for instance if we consider a circular table the spatial invariant distribution is not uniform, being peaked at the boundary \cite{AZ1}. This closely reminds of what is observed in active matter confined in bounded region \cite{algae,PolinNat,DiLnew} (see also \cite{Razin} and references therein): even if active matter interaction with the boundary, causing accumulation, is complex (see for instance \cite{Hyd}), the mechanism of active particles captured by the surface, and released by tumbling might be modelled exactly by (\ref{pEvans}) \cite{MotMic}.\\
The main point of the present paper is to consider small stochastic perturbations of the deterministic elastic reflection law (\ref{el-coll}) \cite{Mark}, and to show that when such a perturbation is very small the billiard tables exhibit either stationary properties associated to Knudsen law (\ref{KnudLamb}) or Evans law (\ref{pEvans}), depending on whether the corresponding deterministic table is chaotic or integrable.\\
First of all let us fix our notation: we consider a particle of unit speed moving ballistically in a bounded region $\Omega \subset  \mathbf{R}^2$  (our billiard table), whose boundary is denoted by $\partial \Omega$: the direction of the velocity is only changed upon collisions with $\partial \Omega$: moreover we parametrise the boundary by the arclength $s\in[0,L)$. For $\mathbf{q} \in \partial \Omega$, $\mathbf{n}_{\mathbf{q}}$ is the unit normal vector directed inwards. Let $\theta_{in} \in (-\pi/2,\pi/2)$ be  the angle (w.r.t. $\mathbf{n}_{\mathbf{q}}$) with which the incoming trajectory would be {\it elastically} reflected upon a collision in $\mathbf{q}$ (the sign is dictated by reference to the oriented tangent to $\partial \Omega$ in $\mathbf{q}$), while $\theta_{out}$ is the actual angle the trajectory takes after collision (measured in the same way as above): so that elastic collisions (deterministic billiard tables) result in $\theta_{in}=\theta_{out}$. Stochastic (Markov) billiards are thus described by a stochastic transition kernel
\begin{equation}
\label{trangle}
{\cal W}(\theta_{in}|\theta_{out}),
\end{equation}
where we are ignore a possible dependence upon the collision point $\mathbf{q}$. With our notation we have 
\begin{equation}
\label{Wel}
{\cal W}(\theta_{in}|\theta_{out})=\delta \left(\theta_{in}-\theta_{out}\right)
\end{equation}
for elastic collisions, and
\begin{equation}
\label{WKE}
{\cal W}(\theta_{in}|\theta_{out})=p_K(\theta_{out}) \quad \mathrm{or} \quad p_E(\theta_{out})
\end{equation}
for Knudsen or Evans stochastic billiards respectively. We will be mainly concerned with the discrete stochastic process $\{\mathbf{q}_n\}$, of consecutive impact points with the boundary, where the particle proceeds with uniform unit velocity with direction induced by $\theta_{out\,(n)}$ after collision at the point $\mathbf{q}_n$, until it hits $\partial \Omega$ again, at $\mathbf{q}_{n+1}$, from which it will exit with velocity directed according to $\theta_{out\,(n+1)}$. If the distribution of outgoing angle is independent of the incoming angle, and given by a common law $p_{\circ}$, like (\ref{WKE}), the relevant process is given by the sequence of collision points: in particular the question is whether it admits an invariant density. We introduce a transition kernel \cite{CPSV} ${\cal K}(\xi,\eta)$, where, $\xi$ and $\eta$ are arclength coordinates corresponding to $\mathbf{x},\,\mathbf{y}\, \in \partial \Omega$
\begin{equation}
\label{Bker}
{\cal K}(\xi,\eta)={\cal K}(\mathbf{x},\mathbf{y})=p_{\circ}(\theta_{\xi})\frac {\cos(\theta_{\eta})}{\Vert \mathbf{x}-\mathbf{y}\Vert}\qquad \mathrm{if}\, \mathbf{x} \rightleftharpoons \mathbf{y},
\end{equation}
where $\theta_{\xi}$ is the angle between $\mathbf{n}_{\mathbf{x}}$ and $(\mathbf{y} -\mathbf{x})$, $\theta_{\eta}$ is the angle between $\mathbf{n}_{\mathbf{y}}$ and $(\mathbf{x} -\mathbf{y})$, while $\mathbf{x} \rightleftharpoons \mathbf{y}$ means that the open segment $(\mathbf{x}-\mathbf{y})$ is fully contained in $\Omega$ (this is always the case for convex billiard tables). Then the invariant density $\mu(s)$ satisfies the equation
\begin{equation}
\label{inv-den}
\mu({\xi})=\int_{\partial \Omega}d\eta\, \mu(\eta){\cal K}(\eta,\xi);
\end{equation}
furthermore if a density $\mu(s)$ satisfies the detailed balance condition 
\begin{equation}
\label{detbal}
\mu(\xi) {\cal K}(\xi,\eta)=\mu(\eta) {\cal K}(\eta,\xi),
\end{equation}
then equation (\ref{inv-den}) is ipso facto satisfied. \\
When we consider the case of Knudsen law (\ref{KnudLamb}), the kernel (\ref{Bker}) becomes {\it symmetric} \cite{CPSV}, so detailed balance is satisfied for a constant density: notice that, in the deterministic case (\ref{el-coll}), when one considers the discrete dynamics on the boundary (with the coordinates arclength $\xi$ and outgoing angle $\psi$) the invariant density
\begin{equation}
\label{bdm}
\mu(\xi)\rho(\psi)=\frac 1 {2|\partial \Omega|} \cos(\psi)
\end{equation}
follows from the uniform invariant measure for the continuous time dynamics \cite{ChMa,Cher,CherFAA}. From a dynamical point of view such a uniform measure coincides with the natural measure (associated to time averages) only when the billiard table is ergodic.\\
On the other side, when we consider a stochastic, uniformly distributed, outgoing angle (\ref{pEvans}) the analysis is not so simple: we do not even know if (apart from few examples, as we will discuss further on) the corresponding Markov chain (\ref{Bker}) satisfies detailed balance \cite{Evans,BA}. \\
We are now ready to formulate our main results: when we introduce a very small stochastic perturbation on the collision dynamics, the billiard becomes ergodic, irrespective of the shape of the table \cite{FY,CPSV,Evans,Fet}, but what about the invariant measure of the
discrete dynamics? We provide analytic arguments and numerical experiments showing that 
for chaotic billiards the invariant density is very close to Knudsen case, while for integrable cases the density strongly resembles Evans billiards: in this sense the stationary probability falls into different universality classes, depending on whether the classical deterministic table is chaotic or integrable. Billiards with mixed phase space require further analysis (see Fig. (\ref{dia-a})).
\\
Before showing our results, we have to define which type of stochastic perturbation we will consider: intuitively we replace the elastic outgoing angle $\theta_{out}^{el}$ with a random variable uniformly distributed in $[\theta_{out}^{el}-\epsilon,\theta_{out}^{el}+\epsilon]$. We cannot use such a definition for any $\theta_{out}^{el}$, since close to $\pm \pi/2$ the trajectory might then escape the table: close to tangency we follow the choice of \cite{Mark}: when $\pi/2-\theta_{out}^{el}<\epsilon$, the outgoing angle is uniformly chosen in $[\pi/2-2\epsilon,\pi/2]$, a symmetric choice being taken around $-\pi/2$. Actually in the final part of the paper we will comment how the choice near tangency is not crucial, unless angular dynamics is absorbing at $\pm \pi/2$ (an example when this may happen is mentioned in \cite{Mark}).\\
Let us first consider billiard tables with ergodic deterministic dynamics: in this case we expect that the Markov chain invariant measure should be close to (\ref{bdm}), due to stochastic stability of strongly chaotic systems \cite{kif,LSY,troub}: this is indeed the case, see Fig. (\ref{dia-a},\ref{dia-s}), when we consider a dispersive diamond billiard \cite{diam,diaM} (the concave region bounded by four arcs of circles of radius $R$ whose centres lie at the vertices of a $L\times L$ square).
\\
It is interesting to check whether the same results hold for billiards with a mixed phase space: we consider here (generalised) lemon billiards 
\cite{postM,ReeR,glem,BZZ,Robn,Loz,robot-lemon,lem-adv-math}: the convex region defined by the intersection of two circles of radii $R_1$ and $R_2$, whose centres are separated by a distance $d$: 
when $R_1=R_2=R$ and $d/R$ is very small the phase space has only a small chaotic component, while for appropriate choices of $R_1 \neq R_2$ and $d$ the billiard is ergodic \cite{glem}.
We see from Fig. (\ref{dia-a}) that indeed in the ergodic case again we recover Knudsen distribution, while strong deviations (manifesting through a flattened distribution) appear close to the integrable case.
\\
When we turn into integrable billiard tables the situation is completely different: here we present results concerning elliptic billiards, the simplest example beyond circular billiards (where rotation symmetry plays an important role). Their purely deterministic dynamics is integrable: any trajectory is tangent to a caustic (confocal ellipse or hyperbola \cite{Tab2,CFell}: see also \cite{Laz}, where the existence of caustics is proven for more general convex tables). What happens if we apply a small stochastic perturbation? We know from \cite{Mark,CPSV} that the system is now ergodic, but as we can see (Fig. (\ref{ell-s},\ref{ell-a})) the spatial and angular density are completely different: they are very well reproduced by Evans results: the outgoing angle distribution (\ref{ell-a}) is flat (apart from a tiny region around the grazing angles $\pm \pi/2$), while the spatial density is not uniform along the boundary, and it is nearly indistinguishable from the invariant spatial distribution of an elliptic Evans billiard. 
\\
\\
As a matter of fact the invariant density for Evans elliptic billiard may be explicitly found \cite{dEv}: let $a$ and $b$ be the semiaxes ($a^{-2}x^2+b^{-2}y^2=1$): then, given two points $\mathbf{q},\,\mathbf{p}$ on the ellipse, we may write the transition kernel (\ref{Bker}) as
\begin{equation}
\label{Eeker}
{\cal K}(\mathbf{q},\mathbf{p})=\frac{1}{\pi} \frac{(\mathbf{q}-\mathbf{p})\cdot \mathbf{n}_{\mathbf{p}}}{\Vert \mathbf{q}-\mathbf{p}\Vert^2}.
\end{equation}
Then, once we write the inward normal at the boundary point $\mathbf{q}$  as
\begin{equation}
\label{ell-nor}
(n_x,n_y)=\frac1{\sqrt{(a^{-4}q_x^2+b^{-4}q_y^2)}}\left(-a^{-2}x,-b^{-2}y\right)
\end{equation}
we can get the identity
\begin{equation}
\label{ell-pro}
\begin{array}{l}
\sqrt{(a^{-4}q_x^2+b^{-4}q_y^2)}\mathbf{n}_{\mathbf{q}}\cdot(\mathbf{p}-\mathbf{q})= \\\sqrt{(a^{-4}p_x^2+b^{-4}p_y^2)}\mathbf{n}_{\mathbf{p}}\cdot(\mathbf{q}-\mathbf{p});
\end{array}
\end{equation}
then some simple calculations show that the invariant density satisfying detailed balance is
\begin{equation}
\label{invEdb}
\rho(\mathbf{q})=\frac C{\sqrt{a^{-4}q_x^2+b^{-4}q_y^2}},
\end{equation}
where $C$ is the normalisation constant \cite{ell-cir}: in Fig. (\ref{ell-s}) we show how (\ref{invEdb}) reproduces with high precision the numerical experiments. We remark that a curvature dependent density on $\partial \Omega$, together with a nonuniform measure in $\Omega$, have been observed in active matter \cite{algae,DiLnew}, and, in this respect, Evans billiard may be considered like a toy model reproducing some key issues of complex matter in confined geometry.\\
As regards the distribution of outgoing angles in the same case, we also get a picture very similar to Evans case: see Fig. (\ref{ell-a}). This is a consequence of the strong correlations of successive outgoing angles in the unperturbed (deterministic) dynamics. Consider for simplicity a circular table: then, in the elastic case with no perturbation, if we start with an outgoing angle $\theta_0$ then the whole temporal sequence will be $\theta_n=\theta_0$: the introduction of the stochastic perturbation amounts then to turn the dynamics into a random walk (we consider
only half of the domain $[0,\pi/2]$, the rule is symmetric in the left half):
\begin{equation}
\label{rwM}
\theta_{i+1}=
\left\{
\begin{array}{ll}
\theta_i+\delta \qquad & \theta_i<\pi/2-\epsilon \\
\pi/2-\epsilon+\delta \qquad & \theta_i>\pi/2-\epsilon
\end{array}
\right. ,
\end{equation}
where $\delta$ is a stochastic variable uniformly distributed in $[-\epsilon,\epsilon]$. For very small perturbations we have a large region $[0,\pi/2-\epsilon]$ where the random walk is homogeneous and symmetric, suggesting a uniform stationary probability distribution, as observed in Fig. (\ref{ell-a}), provided the endpoint $\pi/2$ is not absorbing, which, with our choice of the stochastic perturbation, is guaranteed by a net drift (if $\theta_n> \pi/2-\epsilon$ then $\langle \theta_{n+1}\rangle=\pi/2-\epsilon< \theta_{n}$. The same conclusion can be obtained by writing a Fokker-Planck equation \cite{Risken} and considering its stationary solution. The actual rule of reflection close to $\pm \pi/2$ does not modify the constant density in the bulk of the interval, unless, of course, the boundaries become absorbing.\\
In conclusion, we have considered the effects of small stochastic perturbations of the elastic reflection law of billiard tables: while stochastic stability leads to the preservation of the cosine law and spatial uniform distribution along the boundary in the case of chaotic billiards, the results for integrable billiard tables are quite different: the statistical behaviour is quite close to Evans random billiards, where the reflection angle is a uniformly distributed random variable on $(-\pi/2,\pi/2)$. Such billiards present a number of peculiar statistical properties: the spatial measure along the boundary depends upon the curvature, the volume measure is not uniform (being concentrated along the boundary \cite{AZ1,DiLnew}), Cauchy invariance principle \cite{Cau,San,Maz08,Cab} is not satisfied \cite{AZ1}. In the future we plan to investigate how this picture is modified when ballistic trajectories are replaced by random walks (a natural choice in the description of active particles): though some preliminary indications suggest that boundary conditions are not crucial for walks with a typical length much smaller than the linear size of the confining box \cite{ZA2}, this remains an open problem.
\begin{acknowledgments}
This work has been partially supported by the National Group of Mathematical Physics (GNFM) of the Italian Institute for High Mathematics (INdAM). M.B. is supported by the Insubria University Grant "Statistical properties of random billiards".
\end{acknowledgments}


\begin{references}
\bibitem{ell-cir} Notice that the density becomes constant in the circle limit ($a=b$).
\end{references}
\end{document}